\documentclass[conference]{IEEEtran}
\IEEEoverridecommandlockouts

\usepackage{fancyhdr}
\usepackage{footnote}
\usepackage[hidelinks]{hyperref}
\usepackage{todonotes}
\usepackage{listings}
\usepackage{xcolor}
\fancypagestyle{pageStyle}{%
    \fancyhf{}

}


\ifCLASSINFOpdf
\usepackage{graphicx}
\else
\fi
\usepackage{url}

\begin{document}

%
\title{Towards Traceability in Data Ecosystems\\using a Bill of Materials Model}
\renewcommand\IEEEkeywordsname{Keywords}

\author{\IEEEauthorblockN{Iain Barclay, Alun Preece, Ian Taylor}
\IEEEauthorblockA{Crime and Security Research Institute,\\Cardiff University,\\Cardiff, UK\\
Email: BarclayIS@cardiff.ac.uk}
\and
\IEEEauthorblockN{Dinesh Verma}
\IEEEauthorblockA{IBM TJ Watson Research Center,\\1110 Kitchawan Road,\\Yorktown Heights,\\NY 10598, USA}
}


%



\maketitle
\thispagestyle{pageStyle}
\pagestyle{fancy}
\renewcommand{\headrulewidth}{0pt} 

\begin{abstract}
Researchers and scientists use aggregations of data from a diverse combination of sources, including partners, open data providers and commercial data suppliers. As the complexity of such data ecosystems increases, and in turn leads to the generation of new reusable assets, it becomes ever more difficult to track data usage, and to maintain a clear view on where data in a system has originated and makes onward contributions. Reliable traceability on data usage is needed for accountability, both in demonstrating the right to use data, and having assurance that the data is as it is claimed to be. Society is demanding more accountability in data-driven and artificial intelligence systems deployed and used commercially and in the public sector. This paper introduces the conceptual design of a model for data traceability based on a Bill of Materials scheme, widely used for supply chain traceability in manufacturing industries, and presents details of the architecture and implementation of a gateway built upon the model. Use of the gateway is illustrated through a case study, which demonstrates how data and artifacts used in an experiment would be defined and instantiated to achieve the desired traceability goals, and how blockchain technology can facilitate accurate recordings of transactions between contributors. 

\end{abstract}




%
\IEEEpeerreviewmaketitle

\section{Introduction}
Scientists and researchers increasingly assemble and use rich data ecosystems\cite{oliveira2019investigations} in their experimentation. As these ecosystems expand in capability and leverage data from a diverse combination of internal sources, partners and third party data suppliers, it is becoming necessary for users and curators of data to have reliable traceability on its origins and uses. This can be important to provide accountability\cite{diakopoulos2016accountability}, such as proving ownership or legitimate usage of the source data, as well as being able to identify quality or supply problems and alert users to problems or to seek redress when things go awry. 

Using a gateway to provide traceability on data used within experiments offers mechanisms for demonstrating where data and assets derived from the data are used, as well as aiding understanding where data contributing to a system has come from. By coupling the traceability trail with distributed ledger or blockchain technology, it is possible to provide a distributed store that can record digital data or events in a way that makes them immutable, non-repudiable and identifiable, thereby leading to a trustworthy record of fact.

Research into manufacturing, agricultural and food industries, where the need for traceability of products and their component parts is well-established, has informed the design and development of a gateway which enables data ecosystems to be described in terms of sub-assemblies of their constituent data components and supporting artifacts, in a Bill of Materials (BoM) format. Artifacts in a BoM might include data licenses, software descriptions and versions, and lists of staff or other human resources involved in producing the outputs. When the system described by the BoM is run, the BoM is instantiated, queried for the locations of data sources and populated with any dynamic values for the data or artifacts of each run, generating a Bill of Lots (BoL). The BoM and BoL together provide a record of the static and dynamic elements of the system for an invocation at a particular point in time. This allows for later inspection of the data and the supporting environment, and provides a means for scientists to trace data and artifact usage through and across experiments - for example, identifying all uses of a particular IOT sensor, all runs using a particular version of a machine learning model, or all uses of data generated by a particular researcher.

A pilot gateway, \textit{dataBoM}, has been developed to allow scientists to describe data ecosystem as a Bill of Materials, containing pipelines of assemblies detailing sets of data sources and artifacts, and to instantiate the BoM into a BoL for each run of an experiment. The dataBoM gateway has been developed using GraphQL\cite{byron2017graphql}, which facilitates the rapid development of cross-platform applications and web services which scientists can use to generate and query BoMs and populate and store BoL records. Integration of the dataBoM gateway with blockchain or distributed ledger technologies can provide dynamic behaviour in data acquisition, as well as providing a permanent audit trail of both the data used and its supporting environment.

The remainder of this paper is structured as follows: Section~\ref{sec:related} discusses the context in which the BoM model for data ecosystem traceability has been derived; the architecture and implementation of the dataBoM gateway is discussed in Section~\ref{sec:design}, with Section~\ref{sec:casestudy} describing a case study illustrating how a scientist could use the pilot gateway to conduct research using data from several sources to identify traffic congestion. Section~\ref{sec:discussion} considers areas for future work.

\section{Requirements}
\label{sec:related}

In manufacturing industries it has been standard practice since the late twentieth century to track product through the life-cycle from its origin as raw materials, through component assembly to finished goods in a store, with the relationships and information flows between suppliers and customers recorded and tracked using supply chain management (SCM) processes\cite{lambert1998supply}. In agri-food industries, traceability through the supply chain is necessary to give visibility from a product on a supermarket shelf, back to the farm and to the batch of foodstuff, as well as to other products in which the same batch has been used.

Describing data ecosystems in terms of the data supply chain provides a mechanism to identify data sources and the assets which contribute to the development of the data components, or which are produced as the results of intermediate processes. As new assets are created and used in other systems - perhaps by other parties - the supply chain mapping can be extended to give traceability on the extended data ecosystem.

A definition for traceability is provided by Opara\cite{opara2003traceability}, as "the collection, documentation, maintenance, and application of information related to all processes in the supply chain in a manner that provides guarantee to the consumer and other stakeholders on the origin, location and life history of a product as well as assisting in crises management in the event of a safety and quality breach."

Further helpful terminology is provided by Kelepouris, Pramatari and Doukidis\cite{kelepouris2007rfid} when discussing the traceability of information in terms of the direction of analysis of the supply chain. \textit{Tracing} is the ability to work backwards from any point in the supply chain to find the origin of a product (i.e., `where-from' relationships)\cite{petroff1991framework}. \textit{Tracking} is the ability to work forwards, finding products made up of given constituents (i.e., `where-used' relationships)\cite{petroff1991framework}. Thus, an effective traceability solution should support both tracing and tracking; providing effectiveness in one direction does not necessary deliver effectiveness in the other\cite{kelepouris2007rfid}.

Jansen-Vullers, van  Dorp,  and  Beulens\cite{jansen2003managing} and van Dorp\cite{van2003traceability} discuss the composition of products in terms of a Bill of Materials (BoM) and a Bill of Lots (BoL). 
The BoM is the list of types of component needed to make a finished item of a certain type, whereas the BoL lists the actual components used to create an instance of the item. In other words, the BoM might specify a sub-assembly to be used, and the BoL would identify which exact batch the sub-assembly used in the building of a particular instance of a product was part of. Furthermore, a BoM can be multi-level, wherein components can be used to create sub-assemblies which are subsequently used in several different product types. 

The notion of using a BoM to identify and record component parts of assets in an IT context is already established, with US Department of Commerce working on the NTIA Software Component Transparency initiative to provide a standardised Software BoM\footnote{\url{https://www.ntia.doc.gov/SoftwareTransparency}} format to detail the sub-components in software applications. The intent is to give visibility on the underlying components used in software applications and processes such that vulnerable out-of-date modules can easily be identified and replaced. Tools such as CycloneDX\footnote{\url{https://cyclonedx.org}}, SPDX\footnote{\url{https://spdx.org"}}, and SWID\footnote{\url{https://www.iso.org/standard/65666.html}}) are defining formats for identifying and tracking such sub-components.

As well as the data and any efforts made to secure its provenance\cite{missier2013w3c},\cite{singh2019decision}, there are many supporting assets which can be considered useful supplementary information when recording the characteristics of a data ecosystem. Hind, et al, describe a document based on a Supplier's Declaration of Conformity\cite{hind2018increasing} as a suitable vehicle for providing an overview of an AI system, detailing the purpose, performance, safety, security, and provenance characteristics of the overall system. At the component level, Gebru et al explore the benefits of developing and maintaining Datasheets for Data\cite{gebru2018datasheets}, which replicates the specification documents that often accompany physical components, and Mitchell et al propose a document format for AI model specifications and benchmarks\cite{mitchell2019model}. Schelter, B{\"o}se, Kirschnick, Klein and Seufert\cite{schelter2017automatically} describe a system to automatically document the parameters of machine learning experiments by extracting and archiving the metadata from the model generation process, which would be appropriate information to store alongside the data used in a system.

Many members of the scientific community are familiar with the use of workflow systems, such as Node-RED\cite{node-red} and Pegasus WMS\cite{deelman2015pegasus}, to define and execute the processes for their experiments. The BoM model proposed herein is intended to augment a workflow by providing a means to add contextual traceability, such that it can be archived, and the supporting conditions retrieved and inspected later. Workflow blocks typically describe a job or a service, and do not allow other contributing artifacts to be described. The proposed BoM model describes a rich set of information per node, which can better represent the data supply chain and associated documents and payloads that are contained at each stage. By maintaining a BoM model alongside a workflow, researchers can capture a record of the data for each run, as well as the supporting artifacts for each run, giving traceability of the data and the circumstances in which it was obtained and used.

Distributed ledger technologies, such as those afforded by blockchain platforms\cite{nakamoto2008bitcoin}, \cite{wood2014ethereum}, provide a means of recording information and transactions between parties who do not have formal trust relationships\cite{tapscott2017blockchain}, such as inter-organisational or commercial data sharing entities. The design of a blockchain system ensures that data written cannot be changed, providing a level of immutability and non-repudiation which is well suited to keeping an auditable record of events and transactions which occur between parties. Furthermore, the use of a public blockchain platform, such as the Ethereum Project\cite{wood2014ethereum}, provides an archival resource which remains in existence long after the resources of a project have been retired.
State-of-the-art blockchain platforms, including the Ethereum Project, allow for the deployment of so-called smart contracts, which can be considered to be ``autonomous agents stored in the blockchain, encoded as part of a creation transaction that introduces a contract to the blockchain"\cite{luu2016making}. Such smart contracts  enable blockchain platforms to facilitate non-repudiable dynamic behaviours alongside their immutable storage capabilities.

\section{A Data Traceability Gateway}
\label{sec:design}
In this section the design and implementation of \textit{dataBoM}, a gateway capable of supporting levels of tracking and tracing appropriate for providing traceability in multi-party decentralised data ecosystems, is described. The solution uses a model based on a Bill of Materials scheme, where data and supporting materials are treated as constituent components of a deployed system, which is instantiated into a unique Bill of Lots each time the deployment is run.

\subsection{Conceptual Model}
The dataBoM gateway employs a BoM model, such that each experiment utilising the system is described in terms of its data supply chain. The BoM consists of a collection of \textit{assemblies}, with each assembly being an aggregation of contributing input components and an output component.

An assembly will typically have at least one data input, and can produce new data as its output. Data output from one assembly can be used as a data input in a subsequent assembly within the current BoM, or used in other systems by being referenced in their BoM. To reflect this, data inputs and outputs are defined as \textit{data sources}.

Assemblies can also contain \textit{artifacts}, which are pertinent software components, ML models, and documentation such as licenses, staff lists, policy documentation, etc.. Including artifacts in assemblies in the BoM definition ensures that each BoL retains a full record of its heritage and dependencies.

An assembly can produce a new artifact as its output; for example, an assembly which described the training of an AI model would produce the trained model as its output. The trained model would then be considered an artifact, which could be used as an input to other assemblies.

\begin{figure}[ht]
\centering
\includegraphics[width=0.45\textwidth]{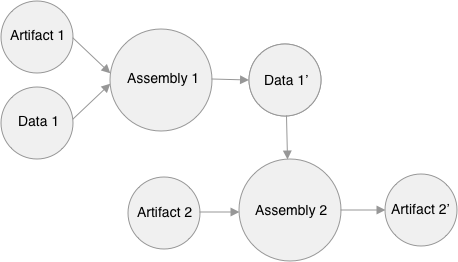}
\caption{Assemblies can be chained in a BoM}
\label{fig:simpleBOM}
\end{figure}

Figure~\ref{fig:simpleBOM} shows two assemblies that are chained to produce a data component (Data 1') and an artifact (Artifact 2') as outputs. Such a BoM could be used by a scientist to describe a simple AI model training process containing two assemblies. Assembly 1 represents the data labelling process, and Assembly 2 the model training process. Data 1 is an input data source, which could be training data. Artifact 1 might be a roster of the staff employed to label the data, and the central data source, Data 1'  (which, as illustrated, is both the output of the data labelling assembly and the input to the model training assembly) could be a labelled data set. In the second assembly, Artifact 2 would be relevant to the model training process, for example the parameters used in training. The output artifact, Artifact 2', would be the trained model. Note that both the intermediate output, Data 1' and the final output, Artifact 2' could be further used as inputs by other processes and specified as inputs to subsequent assemblies. 

The BoM defines a map of the structure of the system by providing a record of the connections between the assemblies, and provides a framework to enumerate a system's data sources and artifacts as well as any static data that applies to the contained data sources or artifacts. This static information could include a location for access to the data (e.g., an API URL), and metadata specifying acceptable data threshold levels or response requirements for active quality of service (QoS) monitoring.

Each time the process described by the BoM is run, the application code for the process will instantiate a new BoL for the given BoM. In order to provide on-going traceability, a shadow data item is created for each data source and artifact in the BoM when it is instantiated in a BoL. The shadow items in the BoL are used to maintain a record of the dynamic elements of each run. 

By storing and then later referencing the assemblies, data sources and artifacts in a BoM, and all the instantiations of the BoM in each BoL, along with the shadow data, it is possible to derive an overview of the history of the data lifecycle of the system, such that any item can be traced back to its origins or tracked forward to find all its consumers.

One of the roles for the data source elements specified in the BoM is to store the means to access the data when the experiment is run. In many cases this will be via a url stored in the data source as part of the BoM, which could be parameterised dynamically at runtime, with the parameters and the results stored in the shadow data item of the BoL. The intent of the design is that there is flexibility of type, so any metadata could be stored in the BoM and retrieved and interpreted in the application process. Uses of this metadata could include storing encrypted information, which is unencrypted and subsequently used by the client application. Further, the metadata could include information to initiate an asynchronous data request and an endpoint to which the data should be delivered. The intent is to provide a flexible storage slot for static data about the data, which can be retrieved, interpreted and used by the client application. Experimentally, it has been possible to use the dataBoM gateway pilot to store and retrieve an encoded blockchain address in a data source, and to initiate a blockchain transaction from the client application to retrieve data at runtime. Such a transaction could be used to provide immutable proof of a data request, or for gateway users to have a means to access third-party data on a pay-per-use basis, which is discussed further in Section~\ref{sec:conclusion}.

\subsection{The dataBoM Gateway}
The dataBoM gateway provides a working implementation of the conceptual data ecosystem BoM model. It enables researchers to declare BoMs to describe the data components of their experiments, and instantiate BoLs to preserve contextual records for each run to provide traceability.

\begin{figure}[ht]
\centering
\includegraphics[width=0.35\textwidth]{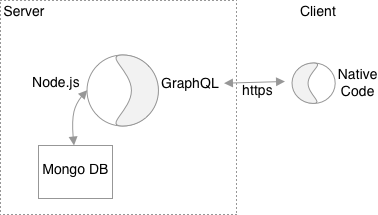}
\caption{The dataBoM Pilot Gateway}
\label{fig:dataBoMArch}
\end{figure}

The architecture of the dataBoM gateway is shown in Figure~\ref{fig:dataBoMArch}. The gateway is offered as a web service, with interactions between the gateway and researchers conducted through a web interface or via an API.

The pilot version of the dataBoM gateway stores data in a MongoDB\footnote{\url{https://www.mongodb.com}} database, such that queries can be written to provide traceability on data sourcing and data use for any BoM. Further development of the gateway will explore the off-loading of the archival of the BoMs and BoLs to commons-based decentralised storage, such as IPFS\cite{benet2014ipfs}, with indexing secured on a public ledger or blockchain. This will serve to preserve records beyond the lifetime of the gateway, and provide an immutable record of events, suitable for later audit or inspection. 

The dataBoM gateway is initially hosted on an intranet, and it is envisaged that future versions of the gateway will be migrated to public facing web services, or serverless\cite{jonas2019cloud} environments, such as AWS Appsync\footnote{\url{https://aws.amazon.com/appsync/}}, to provide a robust and reliable service.

The gateway server is written in Node.js\footnote{\url{https://nodejs.org/en/}}, using Apollo GraphQL Server\footnote{\url{https://www.apollographql.com/docs/apollo-server/}}, which acts as an abstraction layer above the gateway's Mongo DB database store. 

GraphQL allows developers to specify a data schema, and define \textit{queries} and \textit{mutations}, which are interfaces to allow reading and writing of the data, respectively. The GraphQL data schema, queries and mutations are public interfaces, which hide the details of the underlying data storage from users of the interfaces. The server's data store does not have to match the GraphQL schema, as the server code which implements the queries and mutations performs the mapping to read and write the correct data to its database. GraphQL is intended to provide an efficient transfer of data between client and server, as queries can be written to request only the data needed. Furthermore, the gateway's API can be enhanced by extending the queries and mutations offered, without implications for existing users.

The GraphQL interface is self-documenting, and can be queried by client application developers to find out the data structures and queries and mutations available to them.

The dataBoM gateway offers access to its GraphQL server via an https end-point for API access.

\subsection{Integration with Client Applications}
To take advantage of the traceability capabilities provided by the dataBoM gateway, scientists should use the supplied API to define a BoM for their experiments, detailing the assemblies, data sources and artifacts required in their processes, passing the desired parameters and retaining the identifiers which are returned by the API calls in order to chain entities together - for example, when creating a data source item, the identifier that is returned should be retained so that it can be used as a parameter when creating an assembly.

Once the BoM is defined, the researcher should instantiate the BoM whenever they run their experiment, and then use the API from their application code to query the experiment's BoM for static factors such as the locations of data assets, with any dynamic state arising during experimentation (eg. data values) being written to the BoL via the API as the experiment progresses. 

Use of the API requires the researcher to integrate a GraphQL client library with their application code or workflow scripts, and support is available for popular web and mobile platforms, including Python, Node.js, iOS and Android.

The steps in the integration would typically include:

\begin{itemize}
\item Define data sources, artifacts and assemblies in BoM
\item Use BoM's ID to instantiate a new BoL for a new run
\item Access data source metadata for data location or endpoint
\item On receipt of data, populate data source shadow in BoL
\end{itemize}

In this way, the BoM and the BoL can combine to generate an evidence trail of the dynamic data values and the static components of the data and supporting artifacts which contributed to each run of an experiment.

Section~\ref{sec:casestudy}, below, describes a case study implementation, to provide further insight and explanation of dataBoM integration and usage.

\section{Case Study}
\label{sec:casestudy}
\begin{figure}[ht]
\centering
\includegraphics[width=0.3\textwidth]{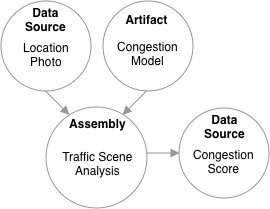}
\caption{The components of a simple traffic congestion system}
\label{fig:HPC}
\end{figure}

By way of illustration of the use of the dataBoM gateway, consider a simple software application which serves to provide a `traffic congestion score' for a fixed location, e.g., Hyde Park Corner, depending on how much traffic the application determines is currently at the location. This simple process has a single assembly, \textit{Traffic Scene Analysis}, an input data source \textit{Location Photo}, an ML model artifact \textit{Congestion Model} and an output data source \textit{Congestion Score} (Figure \ref{fig:HPC}).

In defining the BoM for the Hyde Park Corner (HPC) congestion rating process, the scientist should give each element a name and an optional description, and declare static elements, such as the URL to be used to retrieve a live photo from the location of interest. Encoding this simple single assembly process as a BoM through the gateways's API gives a data model as shown in Listing 1, which is the result of a GraphQL query on the BoM's entry.

\begin{lstlisting}[frame=single,caption={GraphQL data schema for HPC Congestion BoM},captionpos=b]
"bom": {
      "name": "HPC Congestion",
      "description": "Determine congestion levels on Hyde Park Corner",
      "assemblies": [
        {
          "name": "Traffic Scene Analysis",
          "description": "Determine congestion at Hyde Park Corner",
          "inputData": [
            {
             	"name": "Traffic Scene",
                "dataAccess": "https://xyz.com/00001.06514.jpg"
            }
          ],
          "outputData": [
            {
              "name": "Result"
            }
          ],
          "inputArtifacts": [
            {
              "name": "Congestion Model"
            }
          ]
        }
      ]
    }
\end{lstlisting}

In the application code for the experiment, the BoM should be instantiated via its identifier to generate a new BoL for the run. As the code runs, it should refer to its BoM (via the instantiated BoL) to get locations for data it needs to access, and write any  dynamic information to its BoL for permanent archival.

In the HPC congestion scoring example, the data source for the traffic scene holds a static URL for a live camera. The scientist's code would retrieve this information through the dataBoM API and access the photo, and (if desired) store a permanent copy of the photo to its own archives, writing a reference to the location of the archived copy to the shadow data item, such that it will be saved as part of the archival of the BoL. The resultant congestion score should also be written to the BoL, by referencing the appropriate data source item.
 
Thus, each data source and artifact in every BoL would have any dynamic values recorded and stored in a database as a persistent record of the run, so that each of the Assemblies in the BoL would have traceable input and output data values which could be accessed at a later date.

\section{Discussion}
\label{sec:discussion}
There are a number of interesting directions in which future development of the dataBoM gateway could be taken. Interaction with the gateway is currently provided by a GraphQL API, which provides good integration with the application code at runtime, however, initial definition of the BoM and its elements would be more intuitive if it were faciliated through a visual UI. Thus, the BoM could be authored using a visual interface via a web browser, with the runtime invocation and interaction with the BoL remaining an API-driven task. There is a similar opportunity to add a visual interface to the overview of each experiment logged by the gateway. Such an interface would provide a means to explore the composition of the data and artifact elements of each experiment, and help to satisfy the traceability goals of the gateway, by providing a convenient means of exploring the nodes in the BoM and each BoL.

Integration of the dataBoM gateway with the workflow manager systems that are popular in the research community will facilitate smoother integration of the gateway into experiment workflows, and help to foster acceptance of the benefits of the BoM model in providing traceability in scientific data-ecosystems.

There is scope to extend and deepen the integration of the gateway and its BoM and BoL models with blockchain technologies, such as the programmable smart contracts provided by the Ethereum blockchain platform. By associating smart contracts with the data sources and artifacts from the BoM model, novel dynamic behaviour in data ecosystems can be explored. Such dynamic behaviours might include runtime selection of the most appropriate data source sets, along with automatic remuneration and sanctioning, based on dynamic measures of data quality. Further development of the dataBoM gateway could provide a means by which scientists are able to share data and artifacts with their peers, and a blockchain platform might underpin this. Related to blockchain integration is motivation to explore traceability on the human side of the experimental process, using Decentralised Identifiers\footnote{\url{https://w3c-ccg.github.io/did-spec/}} (DIDs) to associate researchers or crowd-workers with components of the system and to provide a means to trace their activity and the data and artifacts they are associated with.

\section{Conclusion}
\label{sec:conclusion}
 The dataBoM gateway provides scientists and developers with a means to map the overall structure of the components that make up complex data ecosystems used in their experiments. By going beyond the data, and considering other contributing factors such as the software and hardware which produces or manages the data, licenses which govern the use and sharing of the data, and policies which contributed to the generation of the data, the development of a BoM for each system provides a mechanism to archive the ecosystem for each experiment. Instantiating the BoM into a BoL each time the system runs augments the static parts list with a dynamic and traceable view into every invocation of the system, such that the data inputs, data outputs and any artifacts which are used or produced by the system can be archived, readily identified and traced back to their source. Similarly, future users of produced data and artifacts, such as models, can be identified, which could prove to be very important if errors are later found and are notifiable. Storing metadata capable of identifying smart contracts on the blockchain further enables immutable recording of the action and timing of requests for data provision, along with the potential for encoding quality of service requirements, and providing automatic payment for services.

\section*{Acknowledgment}

This research was sponsored by the U.S. Army Research Laboratory and the UK Ministry of Defence under Agreement Number W911NF-16-3-0001. The views and conclusions contained in this document are those of the authors and should not be interpreted as representing the official policies, either expressed or implied, of the U.S. Army Research Laboratory, the U.S. Government, the UK Ministry of Defence or the UK Government. The U.S. and UK Governments are authorized to reproduce and distribute reprints for Government purposes notwithstanding any copyright notation hereon.



\bibliographystyle{IEEEtran}
\bibliography{IEEEabrv,datasharing}

\end{document}